\documentclass[sigconf,screen]{acmart}

\usepackage{acronym}
\usepackage{booktabs} 
\usepackage{ccicons}  
\usepackage{todonotes}
\usepackage{subfig}
\graphicspath{{figures/}}
\usepackage[english]{babel}
\usepackage[autostyle=true,english=american]{csquotes}
\usepackage{gensymb}
\usepackage{balance}
\usepackage{xspace}

\DeclareCaptionType{equ}[Equation][]

\newacro{AI}[AI]{Artificial Intelligence}
\newacro{UI}[UI]{user interface}
\newacro{GUI}[GUI]{graphical user interface}
\newacro{TLX}[TLX]{NASA-Task Load Index}
\newacro{RTLX}[NASA RTLX]{NASA Raw-Task Load Index}
\newacro{ER}[ER]{error rate}
\newacro{TCT}[TCT]{task completion time}
\newacro{HCI}[HCI]{Human-Computer Interaction}
\newacro{UX}[UX]{user experience}
\newacro{HFE}[HFE]{Human Factors and Ergonomics}
\newacro{cuDNN}[cuDNN]{ CUDA Deep Neural Network library}
\newacro{RMSE}[RMSE]{root mean squared error}
\newacro{HMD}[HMD]{Head-Mounted Display}
\newacro{RF}[RF]{Random Forest}
\newacro{GP}[GP]{Gaussian process, long-plural = Gaussian processes}
\newacro{KNN}[\textit{k}NN]{\textit{k}-nearest neighbor}
\newacro{NN}[NN]{Neural Network}
\newacro{DNN}[DNN]{ Deep Neural Network}
\newacro{CNN}[CNN]{Convolutional Neural Network}
\newacro{FCL}[FCL]{fully connected layer}
\newacro{BoD}[BoD]{Back-of-Device}
\newacro{FOV}[FoV]{field of view}
\newacro{RW}[RW]{Real World}
\newacro{IFRC}[IFRC]{index finger ray cast}
\newacro{FRC}[FRC]{forearm ray cast}
\newacro{EFRC}[EFRC]{eye-finger ray cast}
\newacro{HRC}[HRC]{Human-Robot Collaboration}
\newacro{HRI}[HRI]{Human-Robot Interaction}
\newacro{6DOF}[6DOF]{six-degree-of-freedom}
\newacro{LOOCV}[LOOCV]{leave-one-out cross-validation}
\newacro{CV}[CV]{cross-validation}
\newacro{RM}[RM]{repeated measure}
\newacro{ANOVA}[ANOVA]{analysis of variance}
\newacro{RMANOVA}[RM-ANOVA]{repeated measures analysis of variance}
\newacro{AGATe}[AGATe]{AGreement Analysis Toolkit}
\newacro{GHoST}[GHoST]{Gesture Heatmap Toolkit Gesture Heatmaps Toolkit}
\newacro{GREAT}[GREAT]{Gesture Relative Accuracy Toolkit}
\newacro{GRT}[GRT]{Gesture Recognition Toolkit}
\newacro{DTW}[DTW]{Dynamic Time Warping}
\newacro{LHRD}[LHRD]{large high resolution display}
\newacro{GEQ}[GEQ]{Game Experience Questionnaire}
\newacro{SPGQ}[SPGQ]{Social Presence Gaming Questionnaire}
\newacro{JND}[JND]{Just-Noticeable Difference}
\newacro{SUS}[SUS]{system usability scale}
\newacro{CSCW}[CSCW]{computer-supported cooperative work}
\newacro{CAD}[CAD]{computer-aided design}
\newacro{MR}[MR]{Mixed Reality}
\newacro{CVE}[CVE]{Collaborative Virtual Environment}
\newacro{AR}[AR]{Augmented Reality}
\newacro{AV}[AV]{Augmented Virtuality}
\newacro{VR}[VR]{Virtual Reality}
\newacro{PRISMA}[PRISMA]{Preferred Reporting Items for Systematic Reviews}
\newacro{PRISMA-Scope}[PRISMA-ScR]{Meta-Analyses Extension for Scoping Reviews}
\newacro{TF-IDF}[TF-IDF]{Term Frequency-Inverse Document Frequency}
\newacro{TF}[TF]{Term Frequency}
\newacro{AVs}[AVs]{Automated Vehicles}
\newacro{eHMIs}[eHMIs]{external Human-machine interfaces}
\newacro{SAR}[SAR]{Spatial Augmented Reality}
\newacro{IFR}[IFR]{International Federation of Robotics}
\newacro{ADLs}[ADLs]{Activities of Daily Living}
\newacro{LED}[LED]{Light-Emitting Diode}
\newacro{DoF}[DoF]{Degrees-of-Freedom}
\newacro{HHC}[HHC]{Human-Human Collaboration}
\newacro{IDF}[IDF]{Inverse Document Frequency}
\newacro{BD}[BD]{Burst Duration}
\newacro{IBI}[IBI]{Inter-Burst Interval}
\newacro{SOA}[SOA]{Inter-Stimulus Onset Asynchrony}
\newacro{ISI}[ISI]{Inter-Stimulus Interval}
\newacro{LRA}[LRA]{Linear Resonant Actuator}
\newacro{TVSS}[TVSS]{Tactile Vision Substitution System}
\newacro{SDK}[SDK]{Software Development Kit}

\AtBeginDocument{%
  \providecommand\BibTeX{{%
    \normalfont B\kern-0.5em{\scshape i\kern-0.25em b}\kern-0.8em\TeX}}}

\copyrightyear{2023}
\acmYear{2023}
\setcopyright{rightsretained}
\acmConference[CHI EA '23]{Extended Abstracts of the 2023 CHI Conference on Human Factors in Computing Systems}{April 23--28, 2023}{Hamburg, Germany}
\acmBooktitle{Extended Abstracts of the 2023 CHI Conference on Human Factors in Computing Systems (CHI EA '23), April 23--28, 2023, Hamburg, Germany}
\acmDOI{10.1145/3544549.3585601}
\acmISBN{978-1-4503-9422-2/23/04}

\acmPrice{15.00}

\newcommand{\conA}{\emph{Rabbit Single}\xspace}
\newcommand{\conB}{\emph{Rabbit Dual}\xspace}
\newcommand{\conC}{\emph{Motion Intensity}\xspace}
\newcommand\change[1]{{#1}}
\begin{document}

\title[HaptiX: Vibrotactile Haptic Feedback for Communication of 3D Directional Cues]{HaptiX: Vibrotactile Haptic Feedback for Communication of \\3D Directional Cues}

\author{Max Pascher}
\orcid{0000-0002-6847-0696}
\email{max.pascher@w-hs.de}
\affiliation{
    \institution{Westphalian University of Applied Sciences}
    \city{Gelsenkirchen}
    \country{Germany}
}
\affiliation{
    \institution{University of Duisburg-Essen}
    \city{Essen}
    \country{Germany}
}

\author{Til Franzen}
\orcid{0000-0003-0203-7512}
\email{til.franzen@studmail.w-hs.de}
\affiliation{
    \institution{Westphalian University of Applied Sciences}
    \city{Gelsenkirchen}
    \country{Germany}
}

\author{Kirill Kronhardt}
\orcid{0000-0002-0460-3787}
\email{kirill.kronhardt@.w-hs.de}
\affiliation{
    \institution{Westphalian University of Applied Sciences}
    \city{Gelsenkirchen}
    \country{Germany}
}

\author{Uwe Gruenefeld}
\orcid{0000-0002-5671-1640}
\email{uwe.gruenefeld@uni-due.de}
\affiliation{
    \institution{University of Duisburg-Essen}
    \city{Essen}
    \country{Germany}
}

\author{Stefan Schneegass}
\orcid{0000-0002-0132-4934}
\email{stefan.schneegass@uni-due.de}
\affiliation{
    \institution{University of Duisburg-Essen}
    \city{Essen}
    \country{Germany}
}

\author{Jens Gerken}
\orcid{0000-0002-0634-3931}
\email{jens.gerken@w-hs.de}
\affiliation{
    \institution{Westphalian University of Applied Sciences}
    \city{Gelsenkirchen}
    \country{Germany}
}

\renewcommand{\shortauthors}{Max Pascher et al.}

\begin{abstract}
In Human-Computer-Interaction, vibrotactile haptic feedback offers the advantage of being independent of \change{any} visual perception of the environment. \change{Most importantly}, the user's field of view is not obscured by user interface elements, and the visual sense is not unnecessarily strained. This is especially advantageous when the visual channel is already busy, or the visual sense is limited. We developed three design variants based on different vibrotactile illusions to communicate 3D directional cues. In particular, we explored two variants based on the vibrotactile illusion of the cutaneous rabbit and one based on apparent vibrotactile  motion. To communicate gradient information, we combined these with pulse-based and intensity-based mapping. A subsequent study showed that the pulse-based variants based on the vibrotactile illusion of the cutaneous rabbit are suitable for communicating both directional and gradient characteristics. The results further show that a representation of 3D directions via vibrations can be effective and \change{beneficial}.
\end{abstract}




\begin{teaserfigure}
\centering
\captionsetup{justification=centering}
    \subfloat[Rabbit Single]{\includegraphics[width=0.32\linewidth]{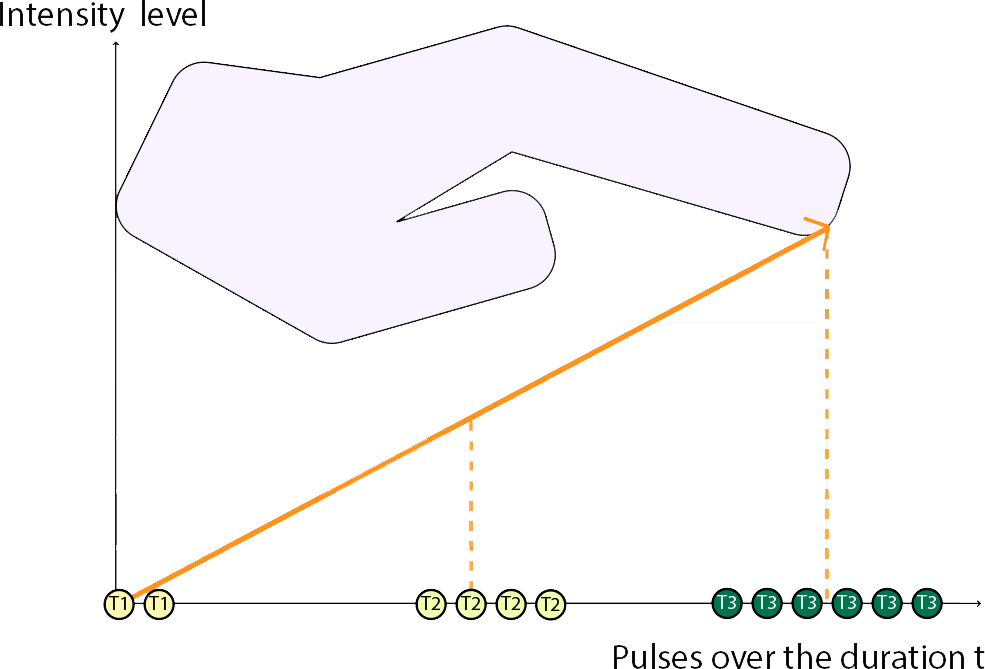}\label{fig:rabbitSingle}}
    \hfill
    \subfloat[Rabbit Dual]{\includegraphics[width=0.32\linewidth]{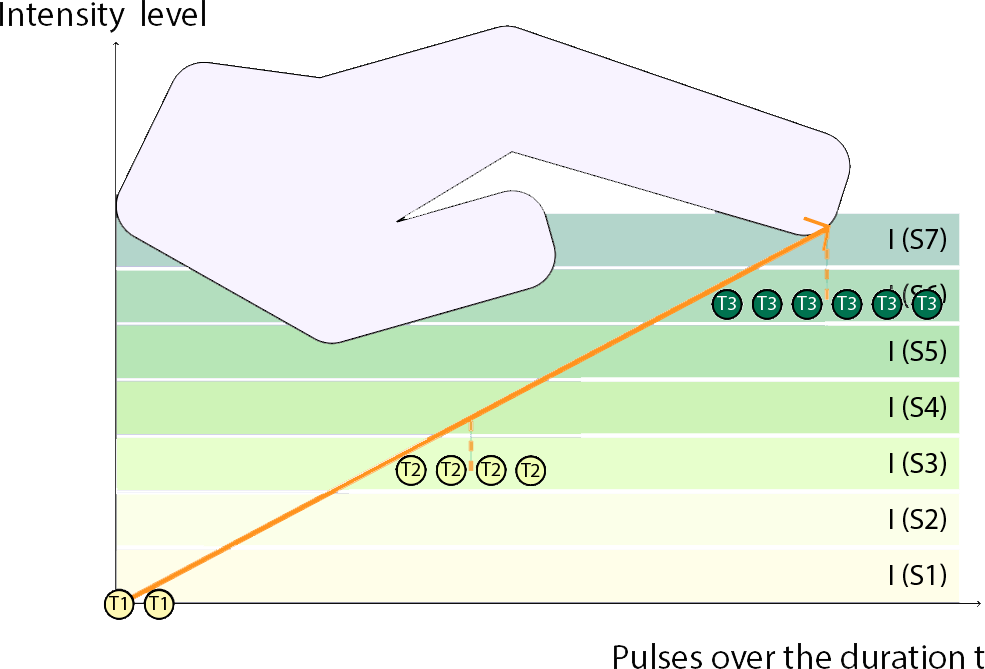}\label{fig:rabbitDual}}
    \hfill
    \subfloat[Motion Intensity]{\includegraphics[width=0.32\linewidth]{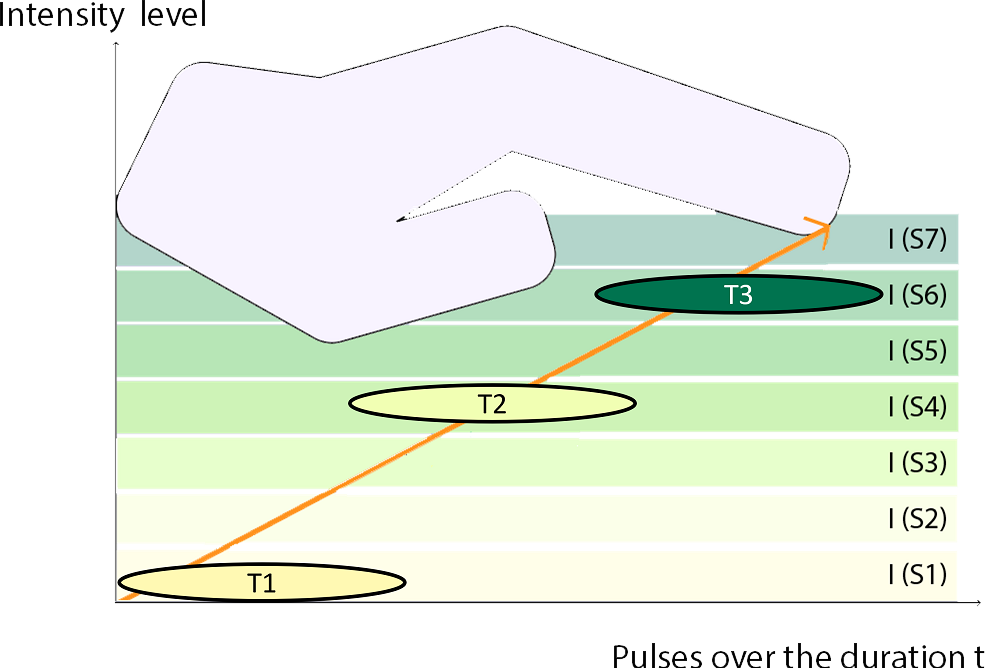}\label{fig:motionIntensity}}
    \hfill
    \captionsetup{justification=justified}
\caption{\change{Coding of the gradient for \textbf{a}) \conA, (\textbf{b}) \conB, and (\textbf{c}) \conC. The orange arrow represents the gradient of the directional cue with an intended increase over time. The timing, duration, number of pulses, and intensity (S1 -- S7) of the three actuators (T1 -- T3) are illustrated for each condition.}}
\Description{An overview of the gradient coding for the three conditions (three figures in a row). All figures consist of a hand, palm facing downward in a diagram t (Pulses over the duration t) over y (Intensity level S1 -- S7). A orange arrow from (0,0) toward the tip of the finger represents the gradient of the directional cue with an intended increase over time. Figure 1a: Rabbit Single: The actuator T1 has two, T2 four, and T3 six pulses, all with the same intensity. Figure 1b: Rabbit Dual: The actuator T1 has two pulses with the intensity S1, T2 four pulses with the intensity S3, and T3 six pulses with the intensity S6. Figure 1c: Motion Intensity: The actuator T1 starts at t = 0 with the intensity S1 and the duration is a third of the whole diagram, overlapping with the actuator T2. The actuator T2 starts with the intensity S4 after a third of the whole time axis in the diagram and the duration is a third of the whole diagram, overlapping with the actuator T3. The actuator T3 starts with the intensity S6 in the last third of the whole time axis in the diagram and the duration is a third of the whole diagram.}
\label{fig:variants}
\end{teaserfigure}

\maketitle

\section{Introduction}
\label{sec:Introduction}

People perceive objects in their environment primarily through their sense of sight. \change{However, this ability can be reduced or not possible at all in certain situations}. \change{Objects might be covered by other things / \ac{UI} elements (visual clutter) or be outside the human field of view}.
In addition, visual perception may be limited or \change{impossible} due to visual impairments.
Previous research has shown that the haptic modality can, to some extent, compensate for the lack of visual information and outperform audio-based cues~\cite{metaanalysis}. It can also be applied in combination with \change{other modalities} and can offer an additional information channel if, for example, the visual channel is overloaded due to distracting information~\cite{Chen2018, Kaul2016HapticHead3G}. 

Directing attention, guiding, and transmitting patterns via vibrotactile signals have already been researched and found to be useful feedback modalities~\cite{Weber.2011,Gunther.2018,Grushko.2021}. \citeauthor{Barralon.2009} studied pattern recognition using a vibrotactile belt with eight actuators and tasked participants to select the corresponding correct visual representation~\cite{Barralon.2009}. \citeauthor{Lee_Starner_2010} proposed \emph{BuzzWear}, a wearable tactile display with three vibration actuators for notification purposes that function by modulating intensity, pattern, direction, and starting point~\cite{Lee_Starner_2010}. After 40 minutes of training, subjects could distinguish between the 24 patterns with up to 99\% accuracy. Vibrotactile feedback is also used in the context of guidance. Here, a study by \citeauthor{Lehtinen2012}, used a vibrotactile glove to support a visual search task on a flat plane on a wall~\cite{Lehtinen2012}.

However, a common challenge is that tactile displays have a limited resolution. Therefore, researchers have simulated smooth movement patterns with the help of tactile illusions~\cite{Cholewiak.2000}, such as \emph{Phantom Sensations}~\cite{4081935,10.1145/3173574.3173832}, \emph{Apparent Tactile Motion}~\cite{Burtt.1917,Kirman.1974,Sherrick.1966}, and \emph{Cutaneous Rabbit}~\cite{Geldard.1977,McDaniel2011,Raisamo2009,10.1145/1518701.1519044}. \citeauthor{Tan2003} conducted a study using a 3 x 3 tactile display and applied the \emph{Cutaneous Rabbit} sensation to explore the communication of eight 2D directional cues (north, northeast, east, southeast, south, southwest, west, and northwest) and the successful recognition of these cues~\cite{Tan2003}.

While previous work focused on 2D directional cues (e.g., \cite{Tan2003}) or allowed users to feel directions upon approach with their hand (e.g., \cite{Grushko.2021}), we are not aware of any work that aims to communicate 3D directional cues. \change{In particular, our work differs from approaches such as~\cite{Gunther.2018}, who aim to push or pull the hand toward a known target in 3D space but who therefore do not actually need to encode 3D information for the vibration pattern itself. It also differs from work such as~\cite{Wu.2013} which used a \ac{TVSS} to communicate 3D shapes of a static object by directly mapping image features such as contours on a 20 x 20 tactile display.} 

Our approach builds on the idea of \citeauthor{Tan2003} \cite{Tan2003} to communicate 2D directions. We combine their base with pulse or intensity mapping to simultaneously communicate the gradient. Furthermore, we explore the influences of different haptic illusions (i.e., \emph{Cutaneous Rabbit} and \emph{Apparent Tactile Motion}) on the comprehension of directional cues. Our work contributes three specific design proposals for communicating 3D directional cues as well as a study on the effectiveness and subjective experience of this non-visual approach to direction mapping.
\section{Concept}
\label{sec:concept}

Within the scope of our experiment, three variants were developed to map vibrotactile 3D directional cues. For the 2D direction, the vibrotactile illusions of the \emph{Cutaneous Rabbit} and \emph{Apparent Tactile Motion} were used. We extended these by a pulse- and intensity-based approach to communicating the gradient of the 3D directional cue (see Figure~\ref{fig:variants}).

\subsection{\conA: Cutaneous Rabbit with Pulse-based Approach}
\label{sec:var-a}
This condition is based on the \emph{Cutaneous Rabbit} for communicating 2D direction, a tactical illusion that can influence the design of vibrotactile patterns. This illusion was discovered in 1972 by \citeauthor{Geldard.1977}~\cite{Geldard.1972}. 
The sequence of taps on different vibrotactile actuators is perceived as a continuous movement between the different points. 
Each directional cue is abstracted using three control points \change{for the} actuators  
\change{(illustrated as dashed lines in \autoref{fig:variants})}. 
Depending on the distance resulting from the gradient of the directional cue, the number of pulses triggered at each actuator is determined in a range of 1 -- 7 with a \ac{BD} of 125ms, an \ac{ISI} between pulses of 50ms, and an \ac{IBI} between actuators of 100ms. 
The closer the control point of the direction cue is to the hand, the higher the number of vibration pulses (see Figure~\ref{fig:rabbitSingle}).

\subsection{\conB: Cutaneous Rabbit with a Pulse- and Intensity-based Combined Approach}
\label{sec:var-b}
\conB is based on \conA but includes a second additional encoding for the gradient of the 3D directional cue. In addition to the number of pulses, we mapped three different intensity levels on the distance of the directional cue to the palm (see Figure~\ref{fig:rabbitDual}). We based the distinct intensity levels on prior work by \citeauthor{Gescheider1990VibrotactileID}, who measured a just noticeable relative difference threshold -- \ac{JND} -- with values of 0.26 at 4 dB above the perceptual threshold~\cite{Gescheider1990VibrotactileID}. To communicate and distinguish between up- and downward gradients, three distinct intensity levels were selected - a baseline level in the middle and one low- as well as one high-intensity level. The anticipated benefit of this condition was that gradient comprehension would be improved due to the dual encoding.

\begin{figure*}
\centering
\captionsetup{justification=centering}
\hfill
    \subfloat[Actuator Placement \change{as provided by \emph{SensorialXR}}]{\includegraphics[width=0.17\linewidth]{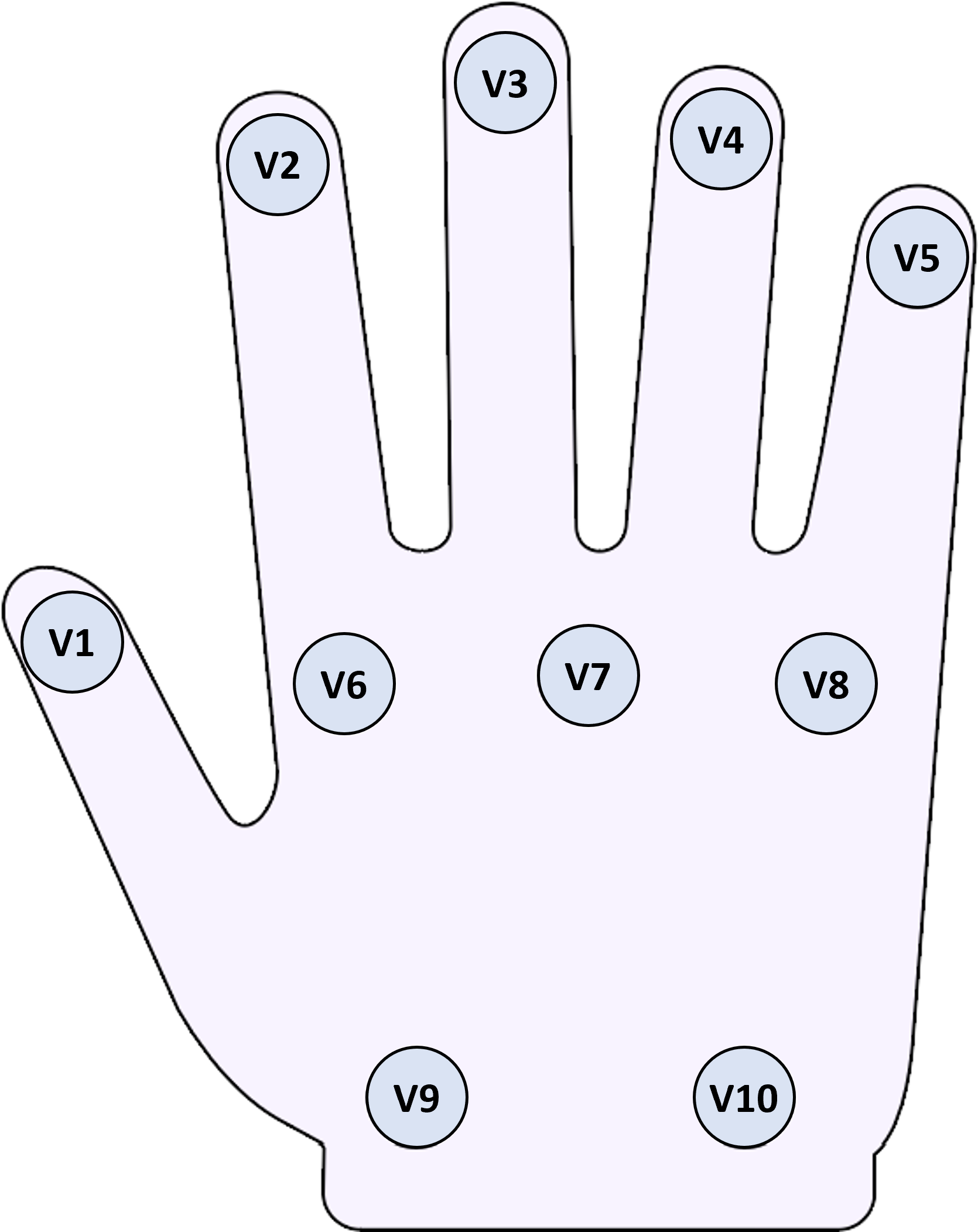}\label{fig:tactors}}
    \hfill
    \subfloat[2D Directions]{\includegraphics[width=0.28\linewidth]{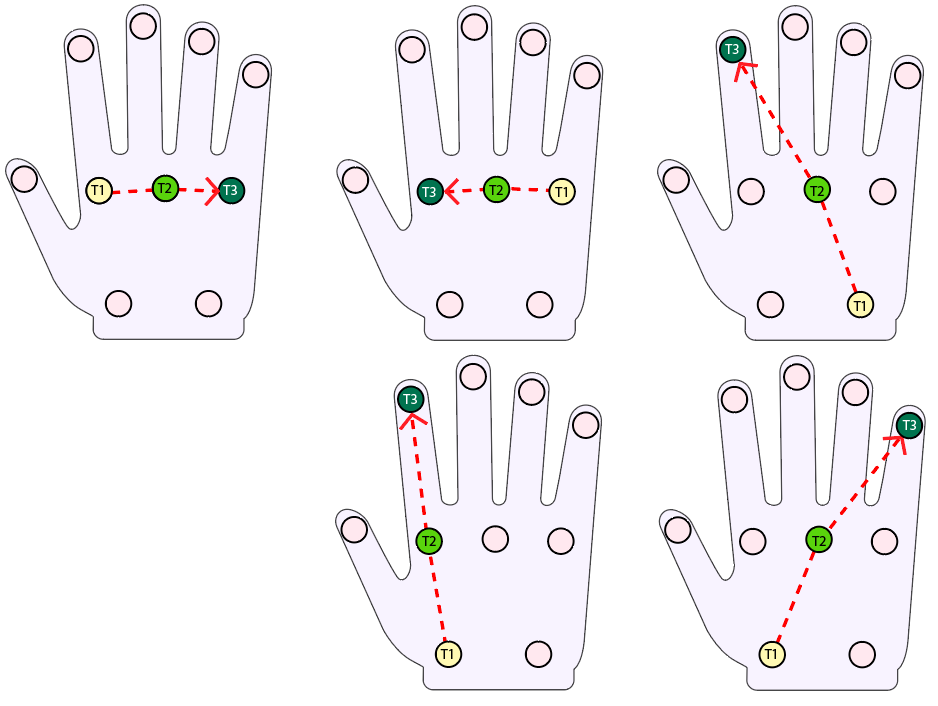}\label{fig:directions}}
    \hfill
    \subfloat[Study Setup]{\includegraphics[width=0.32\linewidth]{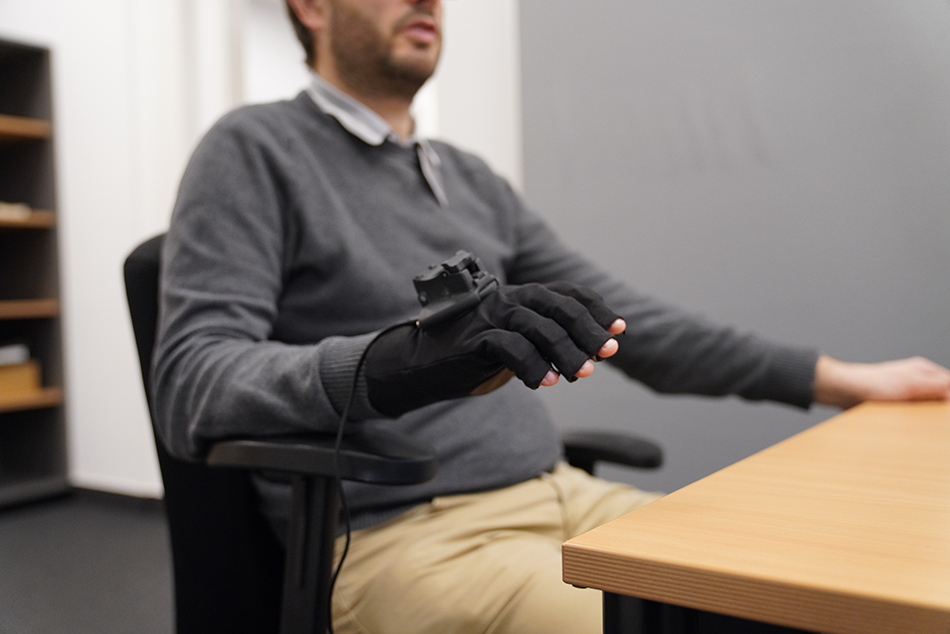}\label{fig:setup}}
    \hfill
\captionsetup{justification=justified}
  \caption{For the 2D directional cues, we used (\textbf{a}) the placement of all actuators across the hand to (\textbf{b}) communicate five different directions. Here, (\textbf{c}) illustrates the study setup, with the arm resting on the armrest while the hand is in the air.}
  \Description{This figures illustrate the actuator placement, 2D directions, and study setup (three figures in a row). Figure 2a: A sketch of a hand with the placement of ten actuators -- at each finger and thumb tip (five; V1 -- V5), between index and middle finger, middle and ring finger, and ring and pinky finger (three; V6 -- V8), and at the left and right bottom of the palm (two; V9 and V10). Figure 2b: Five sketches of the 2D directions with the chronologically activation of three actuators T1 -- T3: left to right (V6 -> V7 -> V8), right to left (V8 -> V7 -> V6), diagonal rear-right to front-left (V10 -> V7 -> V2), straight forward (V9 -> V6 -> V2), and diagonal rear-left to front-right (V9 -> V7 -> V5). Figure 2c: Study setup: Participant is sitting on a chair in front of a table, wearing the glove on the right hand, resting the arm on the chair's armrest and facing the palm downward.}
  \label{fig:hand-setting}
\end{figure*}

\subsection{\conC: Apparent Tactile Motion with Intensity-based Approach}
\label{sec:var-c}
This condition applied the same intensity mapping for the gradient as \conB, but without the pulses. In contrast to the \emph{Cutaneous Rabbit} sensation with distinct pulses as in \conA and \conB, here we applied the vibrotactile illusion of \emph{Apparent Tactile Motion}. This was first studied in the early 20th century by \citeauthor{Burtt.1917}~\cite{Burtt.1917} and is commonly referred to as the \emph{Phi Phenomenon}. The illusion is created by an overlap in the start times of two actuators -- \ac{SOA}, calculated as $SOA = 0.32d + 47.3ms$, where $d$ is the vibration period of an actuator -- 450ms. Instead of two individual actuators, a single stimulus is perceived as moving from the position of the first triggered actuator T1 to the second actuator T2 -- or from actuator T2 to T3 (see Figure~\ref{fig:motionIntensity}). A potential benefit of this illusion is that it may feel more like a natural movement, as it disguises the limited number of actuators. 
\change{After pilot tests, a starting intensity value of 0.22 and a JND value of 0.3 were chosen, which made the intensity levels easily distinguishable. Thus, a total of seven possible intensity levels were defined.}

\subsection{Implementation}
\label{sec:implementation}

To develop our approach, we use the 3D game engine \emph{Unreal Engine 4} optimized for usage with a \emph{Meta Quest~2} \ac{VR} \ac{HMD}. This allows for the use of a virtual environment in which the participants can concentrate purely on the haptic feedback without being visually distracted. It also provides a simple way to visually explain the directional cues to the participants and record their responses for rating scales. As a haptic display, we chose the \emph{SensorialXR} glove as a commercially available device with \ac{SDK} interface to the \emph{Unreal Engine 4}. With ten actuators -- \ac{LRA} vibration motors\change{, fixed in place} -- (see Figure~\ref{fig:tactors}), \emph{SensorialXR} gloves are among the models with the most vibration motors per hand. Thus, they offer the potential to map the 3D directional cues with the highest possible vibrotactile resolution~\cite{Rakkolainen2021}.

\section{Study}
\label{sec:study}
We conducted a within-subjects experiment with 14 participants to explore and understand the differences and similarities between the three presented designs for vibrotactile feedback (independent variable) regarding their effectiveness in communicating 3D directional cues. As participants were supposed to feel and comprehend directional cues without any additional visual feedback, we conducted the study in person and within a neutral VR environment, which allowed participants to focus entirely on the vibrotactile feedback. The age of participants ranged from 21 to 31 years, with a mean age of 25.71 years ($M = 25.71, SD = 2.972$). Four were female, ten were male, and all were university  students of various subjects. None of the participants reported any visual impairment, and all were right-handed.

\subsection{Procedure}
The study was conducted in multiple comparable physical localities. Before commencing, participants were fully informed about the project objective and the various tasks they had to complete. 
Each participant gave their full and informed consent to partake in the study, have video and audio recordings taken, and have all the relevant data documented. 
Participants wore a \ac{HMD} on their head and a vibrotactile glove on the right hand while \change{being asked to keep} their right arm rested on an armrest with the palm facing down (see Figure~\ref{fig:setup}) \change{to avoid any external factors}. 
In the left hand, participants held a controller to control the \ac{VR} environment. 

For each condition, each participant performed six training trials. For each trial, the vibrotactile feedback was repeated three times, and a corresponding visualization was shown to indicate the direction in 3D \change{supporting the participant's mental model}. 
For the actual task, \change{participants were shown a neutral-colored background in VR without any visual representations of the 3D direction. P}articipants were able to trigger the start of the trial with the \ac{VR} motion controller. In total, they completed 30 measured trials per condition, resulting in 90 measured trials per participant and 1,260 measured trials in total. The 30 trials consisted of 2 (blocks) x 5 (2D direction) x 3 (gradient). The variable \emph{2D direction} represented a typical set of five possible mappings of straight horizontal, vertical and diagonal directions, which were physically located on the surface of the hand (see Figure~\ref{fig:directions}). They represented the direction in x-z-coordinates of the overall 3D directional vector. The \change{\emph{gradient}} encoded the direction in y-coordinates: either up, down, or neither any gradient. To counter learning and fatigue effects, we applied a \emph{Balanced Latin Square} design for the order of the three conditions. The order of trials was randomized within each block. \change{Between each condition, participants were able to rest their hand for five minutes.} The average session lasted for 45 minutes and concluded with a debriefing. \change{Non of the participants mentioned any sensory or muscle fatigue.} Participants received 15 EUR in compensation.

\begin{figure*}
    \centering
    \captionsetup{justification=centering}
        \subfloat[2D Direction Estimation.]{\includegraphics[width=0.32\linewidth]{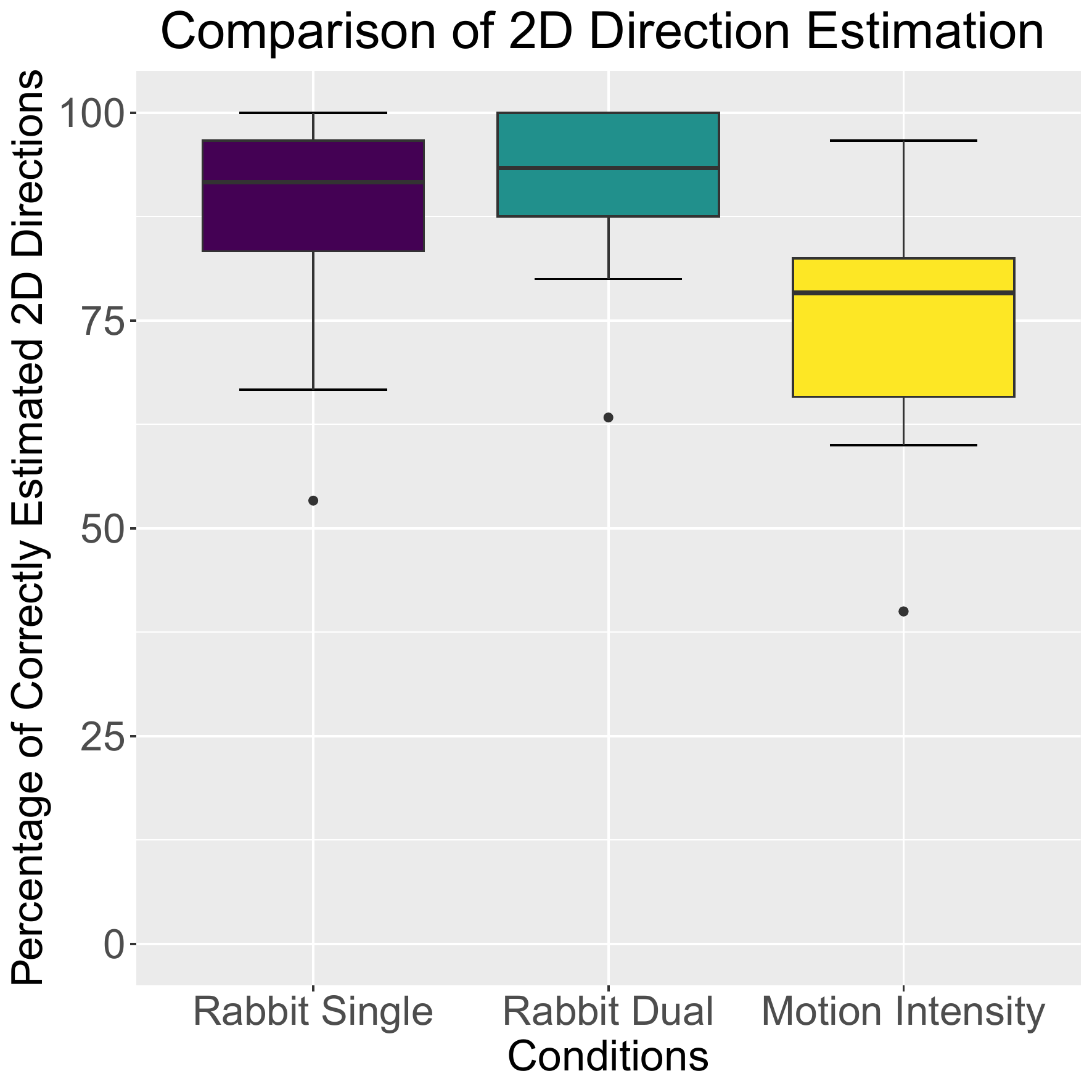}
        \label{fig:boxplots:direction}}
            \hfill
        \subfloat[Gradient Estimation.]{\includegraphics[width=0.32\linewidth]{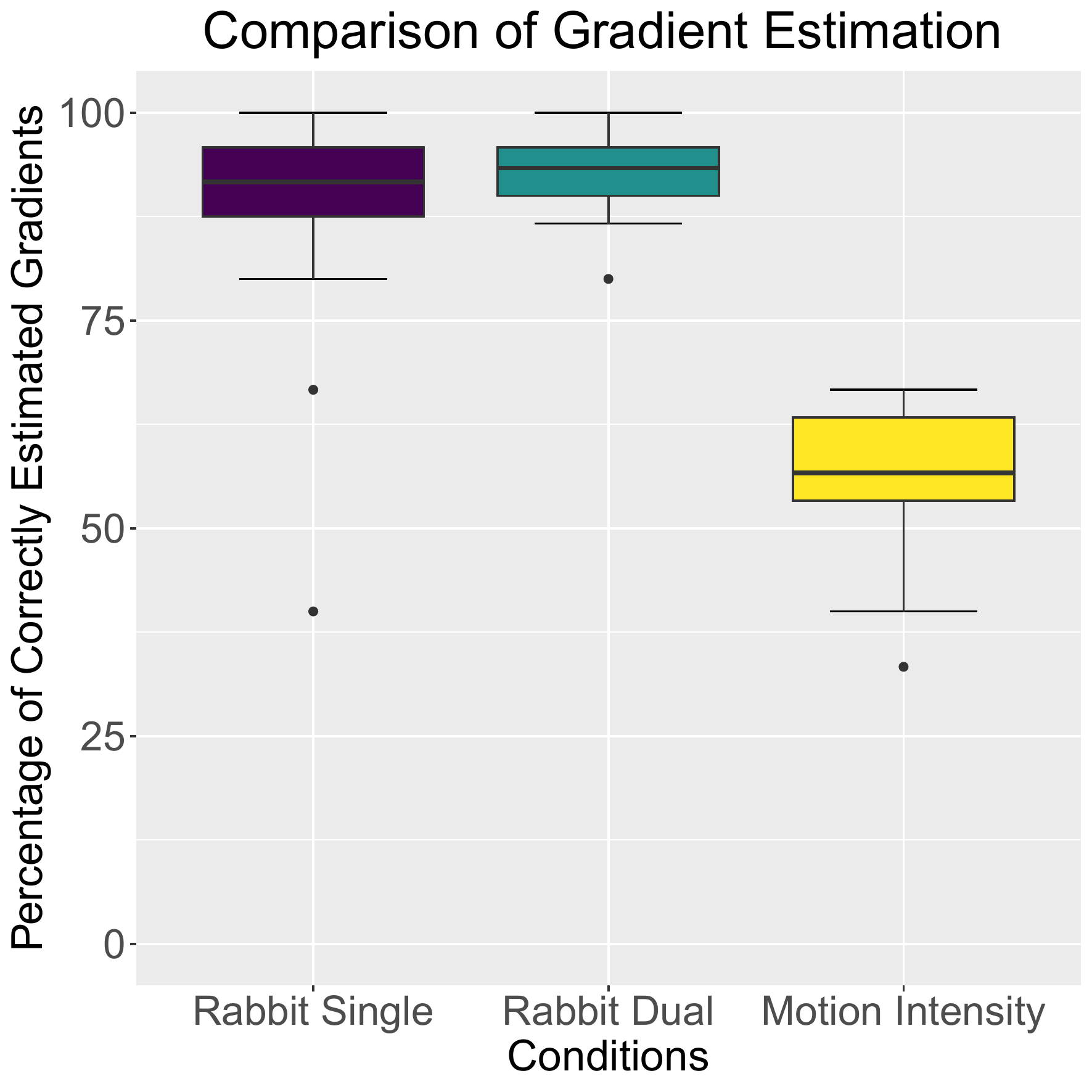}
        \label{fig:boxplots:gradient}}
            \hfill
        \subfloat[Task Load.]{\includegraphics[width=0.32\linewidth]{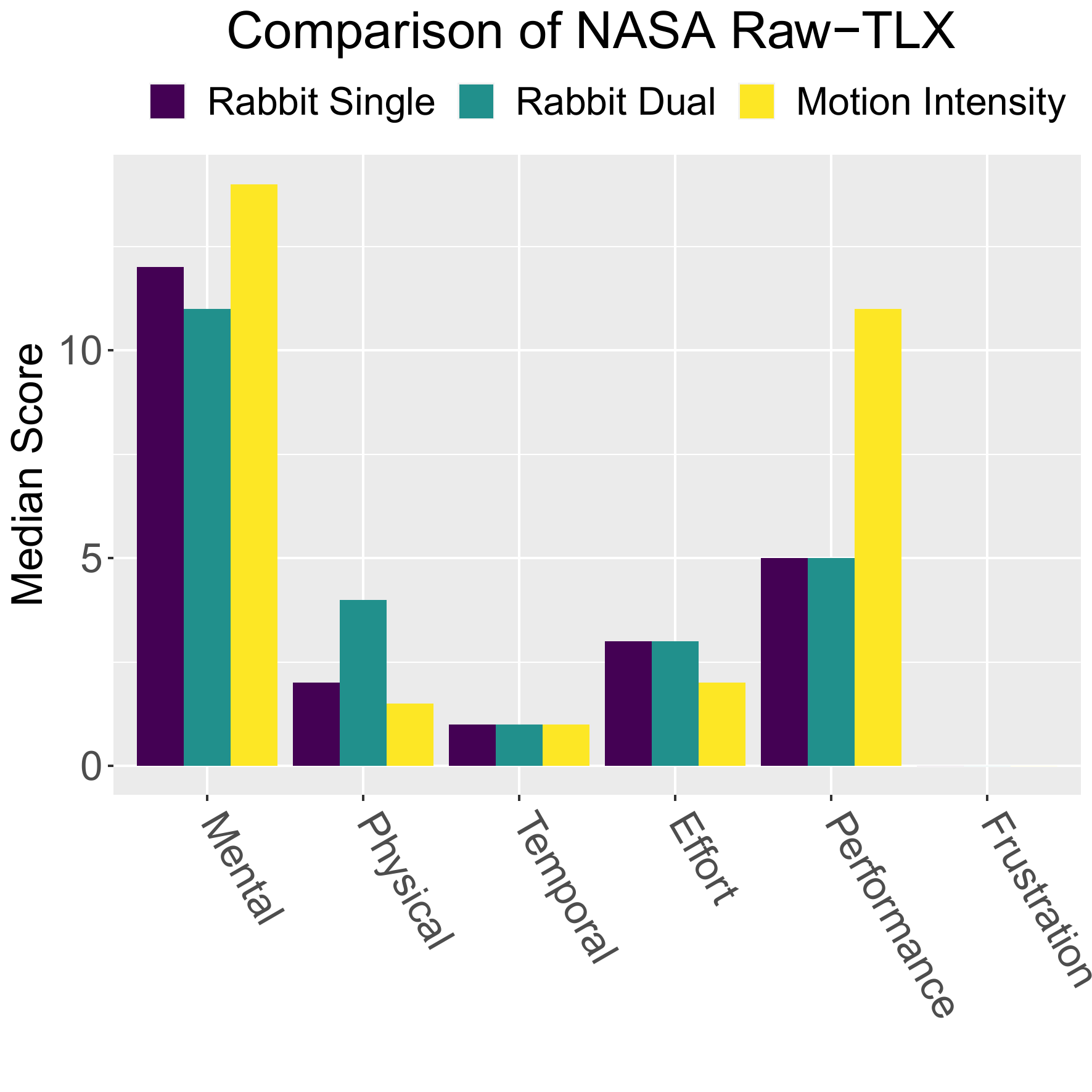}
        \label{fig:boxplots:taskload}}
            \hfill
    \captionsetup{justification=justified}
    \vspace{-0.5em}
    \caption{Measured performance for 2D direction and gradient estimation as well as task load measured with the NASA Raw-TLX (lower score is better). For the task load subscale \enquote{frustration} no bars are visible because all three conditions have a median score of 0.}
    \Description{Three figures in a row, illustrating the results of "Comparison of 2D Direction Estimation" (boxplot), "Comparison of Gradient Estimation" (boxplot), and "Comparison of NASA Raw-TLX" (bar graph) for each condition. Figure 3a: Comparison of 2D Direction Estimation (boxplot): Rabbit Dual=93.3\% (IQR=12.5\%), Rabbit Single=91.7\% (IQR=13.3\%), and Motion Intensity=78.3\% (IQR=16.7\%). Figure 3b: Comparison of Gradient Estimation (boxplot): Rabbit Dual=93.3\% (IQR=5.8\%), Rabbit Single=91.7\% (IQR=8.3\%), and Motion Intensity=56.7\% (IQR=10.0\%). Figure 3c: Comparison of NASA Raw-TLX for the sub-scores Mental, Physical, Temporal, Effort, Performance, and Frustration (bar graph).}
    \label{fig:boxplots}
\end{figure*}

\subsection{Variables and Research Questions}
For dependent variables, we measured the accuracy of the comprehension of the \emph{2D direction} (x-axis, z-axis) and the \emph{gradient} (y-axis). \change{We are measuring the two variables (\emph{2D direction} and \emph{gradient}) separately, as commonly done within the research community (e.g., estimation of direction and distance for \ac{HMD}s~\cite{Gruenefeld2017}). The main reasoning here is that orientation in 3D space and especially describing directions in 3D can be challenging for participants and could negatively affect the validity of the measurements.} To do so, we presented participants with a \ac{UI} panel in VR after each trial. The panel showed five pictures with all 2D directions in a top-down view and, subsequently, three pictures of all gradients in a lateral view. Participants used the \ac{VR} controller to select the fitting representation for each. These two variables were measured with a binary outcome (correct, incorrect) and summarized as percentages of correctly identified outcomes across all trials per person and condition (ratio scale). In addition, we measured mental workload after each condition via the \ac{RTLX}~\cite{hart1988} and additional Likert-scale statements regarding the comprehensibility of the directional cue. We also collected qualitative feedback in a semi-structured interview after each condition as well as at the end of the study, which is when we also asked participants to rank the three conditions. 

As the study is exploratory in nature, we were interested in finding out more about the specific features of our three feedback conditions. In particular, we were interested in the following research questions:

\textbf{RQ1}: Do multiple encodings of gradient, as in the condition \conB, improve the comprehension of gradient information and reduce mental workload? 

\textbf{RQ2}: Does apparent movement, as in the condition \conC, improve comprehension of 2D direction\change{? We assume that to be the case} as the transition \change{by overlapping} of vibration between the actuators \change{may be easier to comprehend and interpret as a path} compared to the sensation of \change{isolated pulses as for the \emph{Rabbit} conditions.} 

\textbf{RQ3}: How would participants experience and rate vibrotactile communication of 3D direction overall and with regard to each individual condition?

\section{Results}
\label{sec:results}

For our applied inferential statistics, we distinguished between ratio and ordinal data. 
The estimation percentages for 2D direction and gradient are ratio data, while the Likert items -- including task load -- are ordinal data.
For ratio data only, we first applied a Shapiro-Wilk test to check for normality.
We found that none of our ratio data is normally distributed.
Thus, we treated all our data in the same way and directly applied non-parametric tests, specifically Friedman tests.
Thereafter, we conducted Wilcoxon Signed-rank tests with Bonferroni correction for our post-hoc analysis.
The effect sizes of the Wilcoxon tests are reported as r (r: $>$0.1 small, $>$0.3 medium, and $>$0.5 large effect).

\subsection{Estimation of 2D Direction}
We asked participants to estimate the two-dimensional direction on a ground plane.
The median (interquartile range) percentages of correct 2D direction estimations for each condition are (in descending order): \conB=93.3\% (IQR=12.5\%), \conA=91.7\% (IQR=13.3\%), and \conC=78.3\% (IQR=16.7\%). 
All percentages are compared in Figure~\ref{fig:boxplots:direction}.
Since our data is not normally distributed (p$<$0.01), we directly ran a Friedman test that revealed a significant effect of condition on 2D direction estimation ($\chi^2$(2)=17.70, p$<$0.001, N=14). 
Post-hoc tests showed significant differences between \conA~and \conC~(W=83, Z=2.62, p$=$0.018, r=0.50) as well as \conB~and \conC~(W=0, Z=-3.30, p$<$0.001, r=0.62).
However, we did not find a significant difference between \conA~and \conB~(W=15, Z=-1.88, p$=$0.182).
Here, \textbf{we can conclude that both \conA and \conB~result in better estimation performance for 2D direction than \conC.}

\begin{table*}
\caption{Pairwise comparisons for individual statements, Bonferroni-adjusted, p-values: $<$0.05 (*), $<$0.01 (**), and $<$0.001 (***).}
\label{tab:pairwise_statements}
\begin{tabular}{l|ll|lll|lll}
 & \multicolumn{2}{l|}{Rabbit Single vs. Dual} & \multicolumn{3}{l|}{Rabbit Single vs. Motion Intensity} & \multicolumn{3}{l}{Rabbit Dual vs. Motion Intensity} \\
Statement & test statistic & p-value & test statistic & p-value & effect size & test statistic & p-value & effect size \\ \hline
S1 & Z=~0.00 & p=1.000 & Z=2.89 & p\textless{}.001\textbf{***} & r=0.55 & Z=3.04 & p=.004\textbf{**} & r=0.57 \\
S2 & Z=-0.07 & p=1.000 & Z=2.58 & p=.029\textbf{*} & r=0.49 & Z=2.80 & p=.013\textbf{*} & r=0.53 \\
S3 & Z=-1.17 & p=0.838 & Z=3.06 & p=.003\textbf{**} & r=0.58 & Z=2.97 & p=.003\textbf{**} & r=0.56 \\
S4 & Z=-0.17 & p=1.000 & Z=2.84 & p=.006\textbf{**} & r=0.54 & Z=2.69 & p=.018\textbf{*} & r=0.51
\end{tabular}
\end{table*}

\subsection{Estimation of Gradient}
We asked participants to estimate the gradient behavior of the communicated cue.
The median (interquartile range) percentages of correct gradient estimations for each condition are (in descending order): \conB=93.3\% (IQR=5.8\%), \conA=91.7\% (IQR=8.3\%), and \conC=56.7\% (IQR=10.0\%). 
All percentages are compared in Figure~\ref{fig:boxplots:gradient}.
Since our data is not normally distributed (p$<$0.001), we ran a Friedman test that revealed a significant effect of condition on gradient estimation ($\chi^2$(2)=19.00, p$<$0.001, N=14). 
Post-hoc tests showed significant differences between \conA~and \conC~(W=102, Z=3.11, p$=$0.002, r=0.59) as well as \conB~and \conC~(W=0, Z=-3.30, p$<$0.001, r=0.62).
However, we did not find a significant difference between \conA~and \conB~(W=30, Z=-1.42, p$=$0.501).
Here, \textbf{we can conclude that both \conA and \conB~result in better gradient estimation performance than \conC.}

\begin{figure*}
    \centering
    \includegraphics[width=\linewidth]{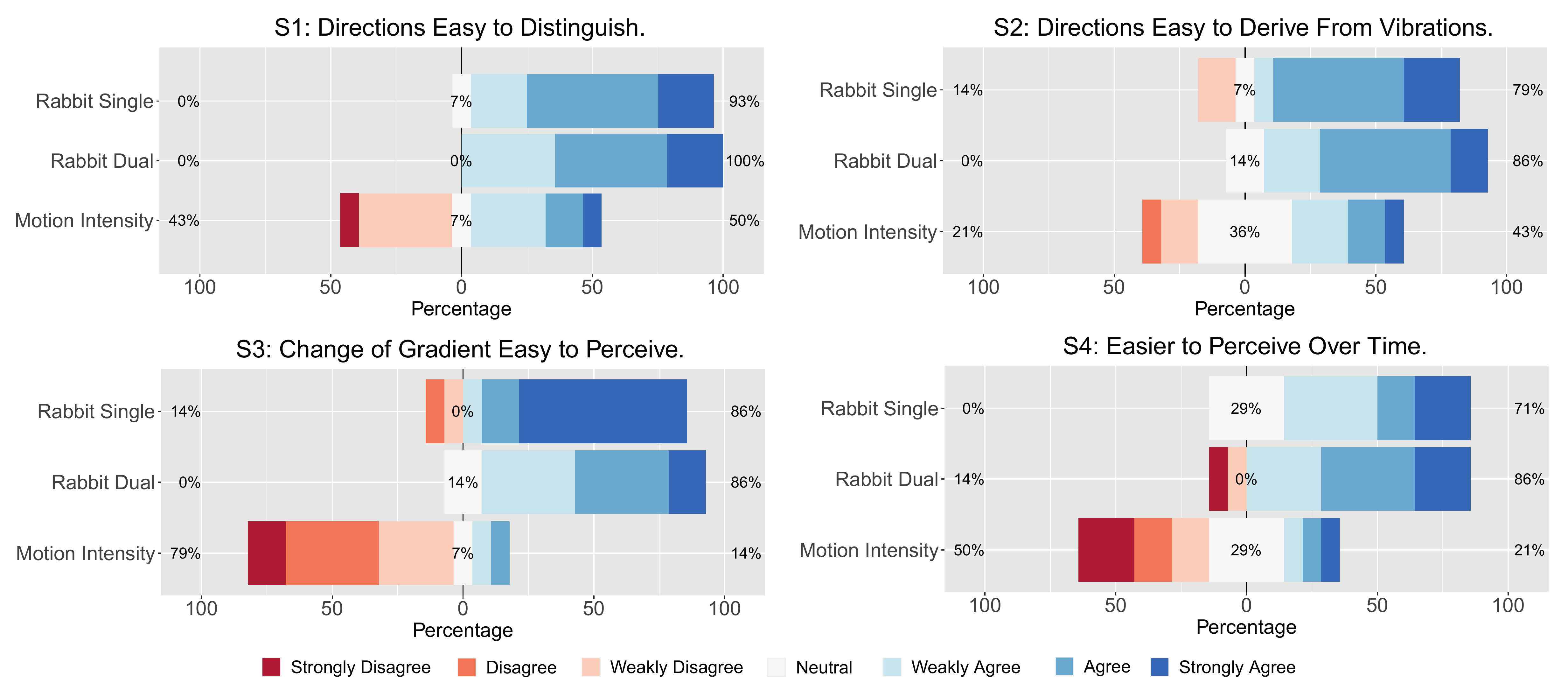}
    \captionsetup{justification=justified}
    \vspace{-1em}
    \caption{Participant responses to the four rated statements (Likert items ranging from 1: strongly disagree to 7: strongly agree).}
    \Description{The results (four figures in two rows with two figures per row) of the participant' statements "Directions Easy to Distinguish", "Directions Easy to Derive From Vibration", "Change of Gradient Easy to Perceive", and "Easier to Perceive Over Time", visualized in a stacked bar plot for Likert items.}
    \label{fig:likert}
\end{figure*}

\subsection{Task Load}
The results of task load ratings as measured by the \ac{RTLX}~\cite{hart1988} are shown in Figure~\ref{fig:boxplots:taskload}. 
The median (interquartile range) task load scores for each condition are (in ascending order): \conA=22.5 (IQR=12.7), \conB=24.5 (IQR=7.9), and \conC=28.3 (IQR=20.0).
We ran a Friedman test that revealed a significant effect of condition on task load ($\chi^2$(2)=13.50, p$=$0.001, N=14).
Post-hoc tests showed a significant difference between \conB~and \conC~(W=105, Z=3.30, p$<$0.001, r=0.62).
However, we did not find any significant differences between \conA~and \conB~(W=39, Z=0.00, p$=$1.000) or between \conA~and \conC~(W=20, Z=-2.04, p=0.120).
Here, \textbf{we can conclude that \conB~induces a lower task load than \conC.}

\subsection{Individual Statements and Preferences}
After each condition, we asked participants to rate four statements, each on a 7-point Likert scale (1=strongly disagree, 7=strongly agree). The results and statements are shown in \autoref{fig:likert}. We found significant main effects for all four statements (N=14; S1: $\chi^2$(2)=14.09, p$<$0.001; S2: $\chi^2$(2)=11.35, p$=$0.003; S3: $\chi^2$(2)=17.08, p$<$0.001; S4:$\chi^2$(2)=12.79, p$=$0.002). Pairwise comparisons are shown in \autoref{tab:pairwise_statements}. Here, \textbf{we can conclude that \conA and \conB~are rated significantly more positively than \conC~for all four statements}. No difference was found between \conA and \conB. Regarding \emph{overall preference}, \textbf{eight participants preferred \conB}, while \textbf{six voted for \conA}~as their favorite. None of the participants preferred \conC.

\subsection{Interviews}
During the interviews, participants were explicitly asked to comment on the duration of the vibration as well as what may have eased or hindered their comprehension. They also had to explain their overall preference and comment on the overall experience and sensation of interpreting 3D directional cues via vibrotactile feedback.
For the analysis, the verbal data was first transcribed by one author and then summarized. The statements were then counted for each question. In addition, across all questions, we applied open coding to identify hidden themes. Data from one interview (P2) was not recorded due to a technical issue. Therefore, only the data from 13 participants was included.

Regarding the duration of the vibration, both \emph{Rabbit} conditions were perceived as having adequate duration (\conA: 10 vs 3 who thought it could have been longer; \conB: 13:0), while 10 participants would have preferred a longer duration for \conC. For the latter, participants struggled to feel the gradient correctly, as mentioned by five participants (e.g., P7 said that the \enquote{[duration was] a little bit short, enough for [2D] direction, but for intensity [gradient] it was really bad.}) The varying strength of the vibration was also an issue, as the most distant control point was criticized as having a too weak vibration, which meant that \enquote{some vibrations got lost} (P5). This also interfered with the comprehension of 2D direction. The smooth transition of movement in \conC was still found to be a pleasant experience, but the mentioned drawbacks regarding the gradient detection prevailed, according to P4 (RQ2). When comparing pulse with intensity for the mapping of gradient, P12 noted an interesting further advantage of pulse, as \enquote{One could decide about the gradient in retrospect even if one wasn't sure before. When the last actuator vibrated many times, then it must have been an upwards gradient.} This also implicitly highlights the problem of immediacy, which required attention and did not allow repetition of the feedback. As P10 put it, \enquote{in case you did not fully pay attention, there wasn't a repeat to make sure.} This sentiment was echoed by P12. 
Consequently, the dual encoding of a gradient in the \conB~condition was cited by most as the main reason for preferring that condition (RQ1). P11 noted, \enquote{I did not just have the number of pulses, but in addition the intensity and that somehow better stuck in my head.}
\section{Discussion}
\label{sec:discussion}
Overall, responses during the interview and the quantitative data are in agreement. They show significant and substantial advantages of \conA and \conB compared to \conC, which we did not expect in such clarity. The parity between these two then is again visible from all angles, with preference being nearly balanced (8 vs. 6). Still, the interviews showed that for \conC, participants did like the smooth transition between the individual feedback factors, which however failed to have a measurable advantage (RQ2). A main reason for this may be that we found that the mechanisms to communicate 2D direction and gradient can interfere with one another. In particular, the intensity gradient mapping had a negative effect on the 2D direction mapping in \conC, as the minimum vibration sometimes \enquote{got lost} (P5), when users did not pay close attention. While pre-tests suggested otherwise, individual differences among the perception of our participants as well as potential fitting issues with the glove (see \emph{limitations} below) may have resulted in this issue. From the comments of participants, we have to assume that \conB was affected by this problem as well, although to a lesser degree; the simultaneous pulse mapping implicitly included repeated vibrations of the same actuator at least twice.

Our analysis also showed that the type of feedback may require more training for participants to get accustomed to. P13 summarized this point nicely: \enquote{Maybe if you market that [...] and someone develops a game for it [...], then I might like it, and in a year, no one can imagine a world without it.} Others noted the effort involved, with P10 saying they \enquote{found that it was really exhausting since you are not used to it.} 
 Still, the overall experience of using vibrotactile feedback to interpret 3D directional cues was described as \enquote{surprisingly good} (P7) and prompted many ideas for use cases, such as medical scenarios (operating table with limited visuals), people with visual impairments in daily life as well as when driving a bike or motorcycle.

\textbf{Limitations:} As an exploratory study, our results should be perceived as preliminary and require further testing and confirmation. In particular, our results may be limited due to the number of participants (14), which also led to the design not being fully balanced. \change{In addition, all participants in our study were right-handed, which could affect our results.} We also found that more training may be required to compensate for initial learning effects, as the type of feedback is so unusual and novel for participants. In addition, the \emph{SensorialXR} glove only \change{provides a fixed setting of the actuators and} offers a \enquote{one-size-fits-all} size, which showed to be problematic for some users with smaller hands, where the actuators were not always in tight contact with the skin. \change{For future research prototypes adding additional Velcro around the actuators might help.} 
\section{Conclusion}
\label{sec:conclusion}

This work aimed to explore different design approaches to communicate three-dimensional directional cues with vibrotactile feedback. We developed two conditions based on the \emph{Cutaneous Rabbit} illusion and one based on \emph{Apparent Tactile Motion} to communicate 2D direction. The gradient of the overall 3D direction was then encoded by the number of discrete vibration pulses, the vibration intensity, or a combination of both.
Our study showed that three-dimensional directional cues can be communicated by \conA and \conB with a high success rate for both the 2D direction and gradient (median for \conA: 91.7\%, \conB: 93.3\%) -- significantly better compared to \conC. With respect to our research questions, we found partial evidence for RQ1, as multiple participants specifically mentioned the dual mapping for gradient as a benefit. Still, for the quantitative data, both \emph{Rabbit} conditions performed more or less identical. RQ2 has to be dismissed at this point. However, as revealed by our qualitative analysis, we believe that the \emph{Apparent Tactile Motion} illusion can also be a viable option for future designs, as the smooth transition between actuators was appreciated by participants. The challenge will lie in overcoming the inferences we found between 2D directional and gradient intensity mapping.

\begin{figure}
    \centering
    \includegraphics[width=\linewidth]{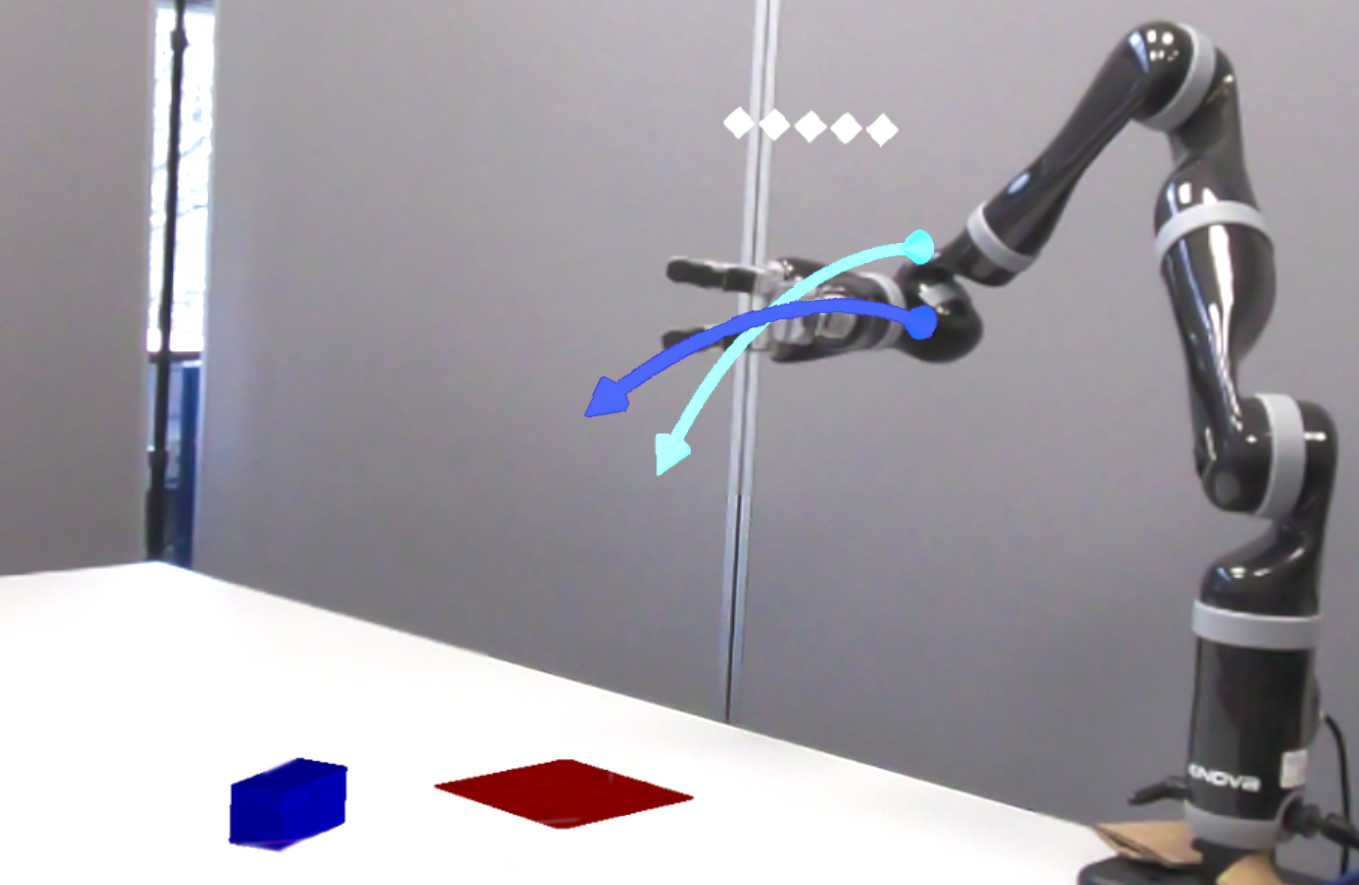}
    \caption{\change{An assistive robotic arm with AI-created directional movement recommendations. The cyan arrow indicate the current movement direction of the arm, while the blue arrow shows the recommendation, which would be mapped as 3D directional cues on the glove. Note: The cyan and blue arrows are only for presentation purposes.}}
    \Description{An overview of a human-robot interaction scenario for future work. A robotic arm (Kinova Jaco) is mounted on a table. The robot is trying to grasp a blue object, and a visual cue (blue arrow) illustrates the intended movement direction toward the object. A cyan arrow points more down toward the table surface, missing the object.}
    \label{fig:future}
\end{figure}

\textbf{Future Research:}  In our work, we aim to apply this approach to communicate the intended movements~\cite{Pascher.2023robotMotionIntent} of a semi-autonomous robot in collaborative scenarios, where vision alone may not be sufficient to successfully predict robot motion. \change{In \autoref{fig:future} an assistive robot arm is illustrated, which is manually controlled by the user but is supported through an \ac{AI} which provides real time directional movement recommendations. Here, our approach could be used to map these directional movement recommendations as vibration input on the hand. Changes in the intensity of the actuators indicate the amount of directional change, thus enabling the user to better imagine the generated trajectory.}
We also encourage researchers to both replicate our design and study and apply it to different use cases. 
Future research should also investigate variables such as the effect of higher-resolution tactile displays\change{, different setting of actuators,} or other approaches to encode gradient (e.g. through different vibration frequencies, \change{varying linear and non-linear intensity levels}), which were not possible with the \emph{SensorialXR} technology.
\change{Furthermore, results of our study should also be evaluated with participants with a dominant left hand or their non-dominant hand.}

\bibliographystyle{ACM-Reference-Format}
\bibliography{bibliography}

\end{document}